\documentclass[reprint,amsmath,amssymb,pre,aps, floatfix]{revtex4-2}

\usepackage{graphicx}
\usepackage{dcolumn}
\usepackage{bm}
\usepackage{xcolor}

\usepackage{times}
\usepackage[english]{babel}

\usepackage{placeins}
\usepackage{hyperref}
\usepackage{natbib}

\begin{document}

\title{Solute dispersion in pre-turbulent confined active nematics}
\author{Tomás Alvim} 
\author{Margarida M. Telo da Gama} 
\author{Rodrigo C. V. Coelho} 
\affiliation{Centro de Física Teórica e Computacional, Faculdade de Ciências, Universidade de Lisboa, 1749-016 Lisboa, Portugal.}
\affiliation{Departamento de Física, Faculdade de Ciências, Universidade de Lisboa, P-1749-016 Lisboa, Portugal.}



\begin{abstract}
  We investigate the dispersion of solutes in active nematic fluids confined in narrow channels based on simulations of nematohydrodynamics. The study focuses on two pre-turbulent regimes: oscillatory flow, with net mass flux, and dancing flow, without net flux. Our findings reveal that the longitudinal dispersion of solutes, in both regimes, is determined by the second moments of the longitudinal and transverse components of the velocity field and proportional to the activity, suggesting a common mechanism for dispersion in these distinct flows. Finally, we found that the dispersion coefficient may increase up to one order of magnitude with respect to molecular diffusion in the dancing flow regime and that it is also enhanced in oscillatory flows, with potential for applications in natural biological environments and engineered devices. 
\end{abstract}
\maketitle


\section{Introduction}
\begin{figure*}
    \centering
    \includegraphics[width=\textwidth]{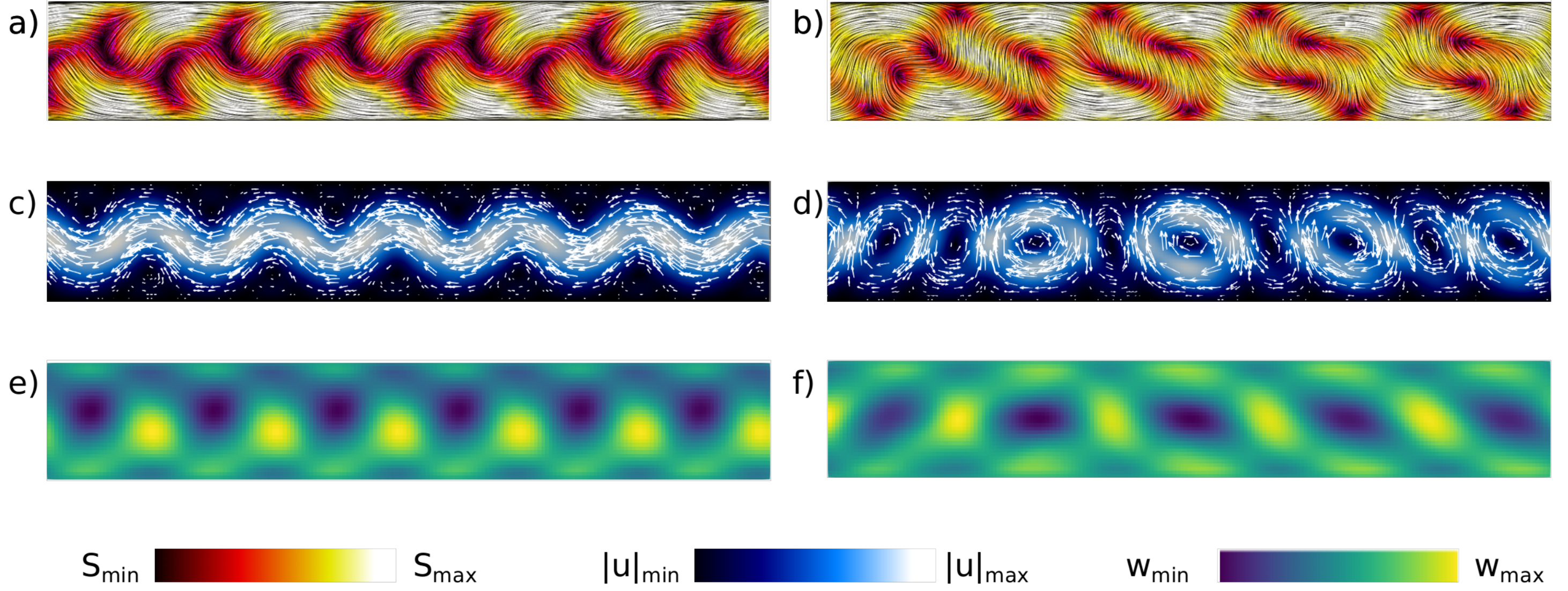}
    
    \caption{ Comparison of the nematic director field, flow velocity, and vorticity in oscillatory (a, c, e) and dancing flow (b, d, f), in a subsection of the channel. The nematic director field is depicted in panels a) and b), visualized using the line integral convolution filter of Paraview~\cite{paraview}. The color map represents the scalar order parameter $S$, which varies from 0 to 0.34 in a) and 0 to 0.42 in b) (see Supplementary Video 1). Velocity and vorticity are given in simulation units. The flow velocity field is shown in panels c) and d), with magnitude reaching $1.8 \times 10^{-3}$ in c) and $3.4 \times 10^{-3}$ in d). The component of the vorticity out of the plane is shown in panels e) and f), varying in the range $-3.9\times10^{-4}$ to $-3.9\times10^{-4}$ in e) and $-1.2\times10^{-3}$ to $9\times10^{-4}$ in f) (see Supplementary Video 2). }
    \label{fig: director flow vorticity}
\end{figure*}

Metabolic and biological processes rely on the consumption and release of molecules, which become dispersed throughout the surrounding fluid. The presence of fluid currents influences the spatial distribution of the dissolved molecules, which governs their local concentrations and thus their effects on biological processes. Fluids confined to channels are a common scenario, both in natural environments, such as porous media like soil \cite{engelhardt_novel_2022}, and in engineered systems, such as lab-on-a-chip devices \cite{kim_microfluidic_2008}. In these settings, fluid flow and the resultant solute transport are strongly affected by the fluid's confinement and state with emphasis on its activity.

A physical model that describes many aspects of the self-generated active flows is nematohydrodynamics with non-equilibrium stresses that couple linearly the activity to the orientational order \cite{marenduzzo_steady-state_2007} as the dominant far-field contribution of a microswimmer to the velocity field typically resembles that of a force dipole, apolar at the microscale. 
Theoretical analysis of tracer dispersion in suspensions of decorrelated squirming microswimmers predicts an enhanced diffusivity proportional to both the self-propulsion velocity and the number density of microswimmers \cite{lin_stirring_2011}. Experimental studies of tracer dispersion in mesoscale turbulence report a dispersion coefficient proportional to the swimmer density \cite{wu_particle_2000,leptos_dynamics_2009}. The density of the swimmers and their self-propulsion are taken into account as a single activity parameter that sets the scale of the active stresses in the hydrodynamic theory \cite{aditi_simha_hydrodynamic_2002}. 

Self-organized flows, generated by swimming microorganisms \cite{leptos_dynamics_2009} or active sub-cellular components like microtubules \cite{wolke_actin-dependent_2007}, are unlike any passive fluid flow. In large systems, these flows tend to become chaotic, in what has been called active turbulence \cite{alert_active_2022}, causing colloidal particles to transition from ballistic to diffusive dispersion due to the loss of spatial and temporal correlations in the flow \cite{wu_particle_2000}. However, when microorganisms or microtubules are confined within a channel whose width matches the characteristic length, such as the mean vortex size, screening of the active fluctuations renders the flows more regular. These flows can be split into directional flows and vortex lattices, the former occurring in narrow and the latter in wider channels for fluids with the same activity \cite{chandragiri_active_2019}. This transition, driven by the channel width, has also been observed experimentally in systems of microtubules \cite{hardouin_reconfigurable_2019} and suspended bacteria \cite{wioland_directed_2016,henshaw_dynamic_2023}.

In previous studies, channel flow was simulated using active nematic models, and is reviewed in \cite{thampi_channel_2022}, from where we borrow the nomenclature. The directional flow is referred to as oscillatory flow due to its periodic undulations. Similarly, the vortex lattice, characterized by dynamic vortices and defect motion, is referred to as dancing flow \cite{shendruk_dancing_2017}. In the dancing flow regime, the periodic motion of the defects forms a spatiotemporal structure known as the silver braid, which has been shown to generate the highest topological entropy for a linear arrangement of defects \cite{tan_topological_2019}. Other confinement geometries have also been shown to enhance mixing efficiency \cite{memarian_controlling_2024}. High entropy states suggest potential applications for efficient micro-scale mixing, where understanding the dispersion behavior, as explored in this paper, is relevant. 

Transitions between pre-turbulent flow regimes are controlled by varying either the width of the channel or the characteristic length of the active flow, defined as $l_a = \sqrt{K/\zeta}$ where $K$ represents the elastic constant penalizing distortions in the orientational order, and $\zeta$ is the activity parameter. The confinement-induced order provides a means to control the flow dynamics and the solute dispersion. Beyond channel confined active nematics, dispersion has been studied in other active fluids and geometries~\cite{hernandez-ortiz_transport_2005, underhill_diffusion_2008, mikhailov_hydrodynamic_2017, dehkharghani_bacterial_2019, bate_self-mixing_2022, modica_boundary_2023, ge_hydrodynamic_2023}.

It has been reported that in confined active layers the velocity fluctuations render the diffusion coefficient of tracers dependent on the system size and divergent as the thickness approaches the spontaneous flow transition \cite{basu_anomalous_2012}.
Somewhat surprisingly, a recent analytical study of solute transport in cylindrical capillaries under activity-gradient driven flows reported lower solute dispersion when compared with the dispersion in passive fluids driven by pressure gradients \cite{das_taylor_2024}.

Despite these studies, the effect of confinement on solute dispersion in active nematics flows remains largely unexplored. Here we aim to start closing this gap.
Specifically, we aim to investigate how solute dispersion differs in self-organized flows with net flux from those without. We hypothesized that the transition from directional to nondirectional flow would result in a discontinuity of the effective diffusivity.
Using lattice Boltzmann simulations of a confined 2D active nematic fluid, we found no such discontinuity for dissolved solutes. Instead, we found a generic relationship
between the effective diffusivity and the second moments of the velocity field through an active form of Taylor dispersion \cite{aris_dispersion_1956}, characterized by the same transverse diffusion length scale in both regimes. These findings provide new insights into the dispersion of nutrients and other molecules in channels containing microswimmers under different flow conditions.

\section{Methods} 
\begin{figure*}
    \centering
    \includegraphics[width=0.8\textwidth]{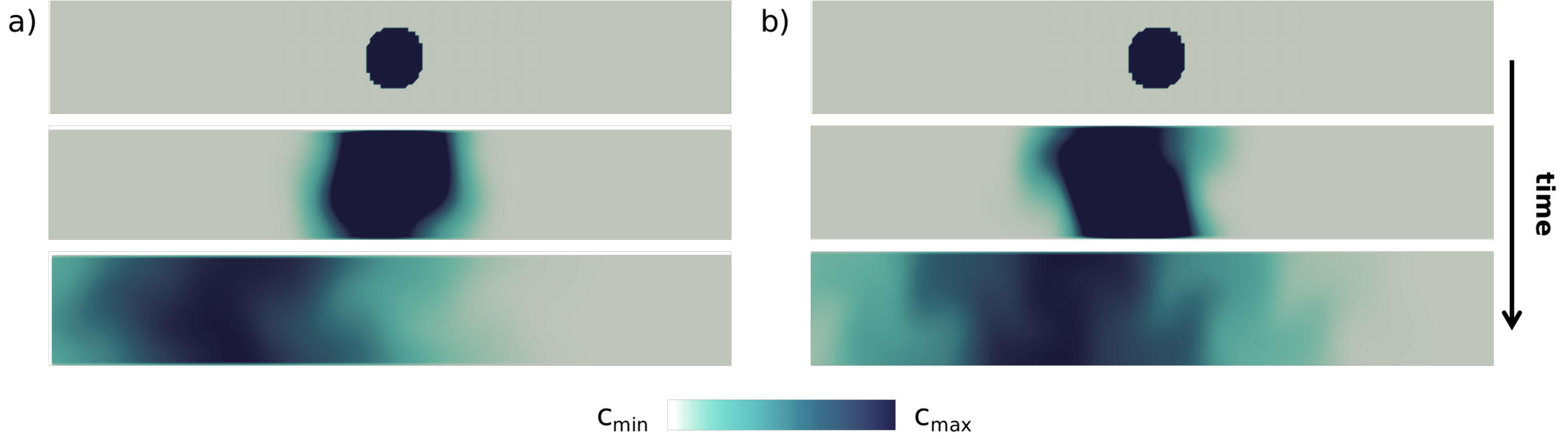}
    
    \caption{ Evolution of a solute drop in oscillatory a) and dancing flow b) regimes, in the section of the channel where the solute drop is initially placed. The solute concentration varies from $0$ to $10^{-1} c_0$ in a), where $c_0$ is the initial concentration. The color scheme is saturated in the initial setup (first row) for visualization purposes. In b) the solute concentration varies from $0$ to $10^{-2} c_0$. }
    \label{fig: soluto}
\end{figure*}

In order to simulate active nematics, we use the Landau-de Gennes free energy, written in terms of the order parameter $\mathbf{Q}$. In two-dimensional uniaxial nematics $\mathbf{Q}= S(\mathbf{n n}-\frac{1}{2}\mathbf{I})$, where $\mathbf{n}$ is the director field, corresponding to the eigenvector with the largest eigenvalue of $\mathbf{Q}$, and represents the average local orientation of the molecules. The degree of orientation is quantified by the scalar order parameter $S$, describing an isotropic fluid at $S=0$ and a perfectly ordered nematic at $S=1$.  To describe the fluid motion and its couplings to nematic orientations, we use the Beris-Edwards theory of nematodynamics, starting with the advection-diffusion equation for $\mathbf{Q}$:
\begin{eqnarray}
    (\partial_t +u_\gamma \partial_\gamma) Q_{\alpha \beta} - S_{\alpha \beta} = \Gamma H_{\alpha \beta},
    \label{eq: beris edward dynamics}
\end{eqnarray}
where $\Gamma$ is the rotational diffusivity. The tensor $S_{\alpha \beta}$ accounts for the rotations of the director due to vorticity and shear strain rate, and it is expressed as:
\begin{equation}
\begin{aligned}
S_{\alpha \beta} &= (\xi D_{\alpha \gamma}+W_{\alpha \gamma})\left(Q_{\beta\gamma}+\frac{\delta_{\beta\gamma}}{2}\right) \\
&+\left(Q_{\alpha \gamma}+\frac{\delta_{\alpha \gamma}}{2}\right)(\xi D_{\gamma\beta}-W_{\gamma\beta}) \\
& -2\xi \left( Q_{\alpha \beta}+\frac{\delta_{\alpha \beta}}{2}\right) (Q_{\gamma\epsilon}\partial_\gamma u_\epsilon),
\end{aligned}
\end{equation}
where $\xi$ is the alignment parameter, which is positive in the flow aligning regime. Here, the vorticity tensor $W_{\alpha \beta}$ is defined as $W_{\alpha \beta}=\frac{1}{2}(\partial_\beta u_\alpha-\partial_\alpha u_\beta)$, and the shear rate tensor $D_{\alpha \beta}$ is defined as $D_{\alpha \beta}=\frac{1}{2}(\partial_\alpha u_\beta+\partial_\beta u_\alpha)$. The molecular field $\mathbf{H}$ :
\begin{eqnarray}
    H_{\alpha \beta} = - \frac{\delta F}{\delta Q_{\alpha \beta}} +\frac{\delta_{\alpha \beta}}{2}\text{Tr}(\frac{\delta F}{\delta Q_{\gamma\epsilon}}),
    \label{eq: molecular field}
\end{eqnarray}
is obtained from the following elastic free energy density:
\begin{equation}
    \begin{aligned}
        f_e = \frac{1}{2}K (\partial_\gamma Q_{\alpha\beta})^2,
    \end{aligned}
\end{equation}
 with elastic constant $K$. The anchoring energy density at the channel walls is:
\begin{equation}
    \begin{aligned}
      f_a = \frac{1}{2}W_0 (Q_{\alpha\beta}^0-Q_{\alpha\beta})^2,
    \end{aligned}
\end{equation}
where $W_0$ is the anchoring strength and the preferred surface order is determined by $\mathbf{Q}^0$. The total free energy is $F=\int (f_e+f_a)d^3r$. The reason for the absence of a bulk free energy density in this model for the active nematic is that many physical systems, such as the microtubule kinesin mixtures, do not exhibit nematic order until they become active \cite{xi_material_2018} \cite{gupta_adaptive_2015}. Furthermore, theoretical and numerical results show that activity induces order in flow-aligning suspensions of force dipoles of extensile systems~\cite{thampi_intrinsic_2015} \cite{santhosh_activity_2020}. The walls of the channel will induce nematic order close to them due to the anchoring energy. \par
We consider an incompressible fluid governed by the Navier-Stokes equation, with the continuity equation $\partial_\beta u_\beta=0$ and the momentum equation:
\begin{eqnarray}
  \rho \frac{D u_\alpha}{D t} = \partial_\beta (\Pi^{p}_{\alpha \beta} +\Pi^{a}_{\alpha \beta})  \label{eq: NS}\\
 \Pi^{a}_{\alpha \beta}=-\zeta Q_{\alpha \beta}\\
 \Pi^{p}_{\alpha \beta} = 2\eta D_{\alpha\beta} - p\delta_{\alpha\beta} ,
\end{eqnarray}
where the superscripts ``p'' and ``a'' denote passive and active.  The activity parameter $\zeta$ quantifies the concentration and magnitude of the microscopic force dipoles, and $\eta$ is the fluid's dynamic viscosity. There are no external pressure gradients or body forces imposed on the system.
\par The dispersed solute follows the diffusion equation for the concentration field c, with molecular diffusivity $D_m$:
\begin{eqnarray}
    (\partial_t +u_\gamma \partial_\gamma)c(\mathbf{r},t)= D_m \partial_\alpha^2 c(\mathbf{r},t).
\end{eqnarray}
\par We simulate this complex fluid in two dimensions, with a combination of two methods. For Eq.~\eqref{eq: NS} the lattice Boltzmann method, incorporating the active stress through the Guo force method \cite{kruger_lattice_2017}. For Eq.~\eqref{eq: beris edward dynamics}, a finite-differences method with predictor-corrector integration \cite{marenduzzo_steady-state_2007}. The channel has two walls separated by a distance of $L=32$. The lattice constant $\Delta x$, integration time step $\Delta t$, and the reference density $\rho_r$ are set to 1, which is known as lattice units, and it will be used throughout the paper. By relating the unit length, time, and density to those values in experiments, the lattice units can be converted to physical units. We use the ratio of the active length to the channel width, to define the activity number $A=L/l_a$, for comparison with previous studies, such as reference \cite{samui_flow_2021}. 

\begin{table}
    \centering
    
     \caption{Simulation parameters in lattice units.}
    \begin{tabular}{|c|c|}
        \hline
        \textbf{parameter} & \textbf{value} \\
        \hline
        viscosity $\nu$ & 1/6 \\
        \hline
        density $\rho$ & 40 \\
        \hline
        single elastic constant $K$ & 0.015 \\
        \hline
        alignment parameter $\xi$ & 0.90 \\
        \hline
        anchoring strength $W_0$  & $2.0 \times 10^{-3}$ \\
        \hline
        molecular diffusivity $D_m$ & $3.33 \times 10^{-3}$ \\
        \hline
        channel dimensions  $L_x \times L$ & $1024 \text{ or } 2048 \times 32$ \\
        \hline
        reference oscillatory activity $\zeta_o$ & $8.5\times 10^{-3}$ \\
        \hline
        reference dancing activity $\zeta_d$ & $1.3\times 10^{-2}$ \\
        \hline
    \end{tabular}
    \label{tab: sim param}
\end{table}

\section{Results}

\begin{figure}
    \centering
        \includegraphics[width=0.4\textwidth]{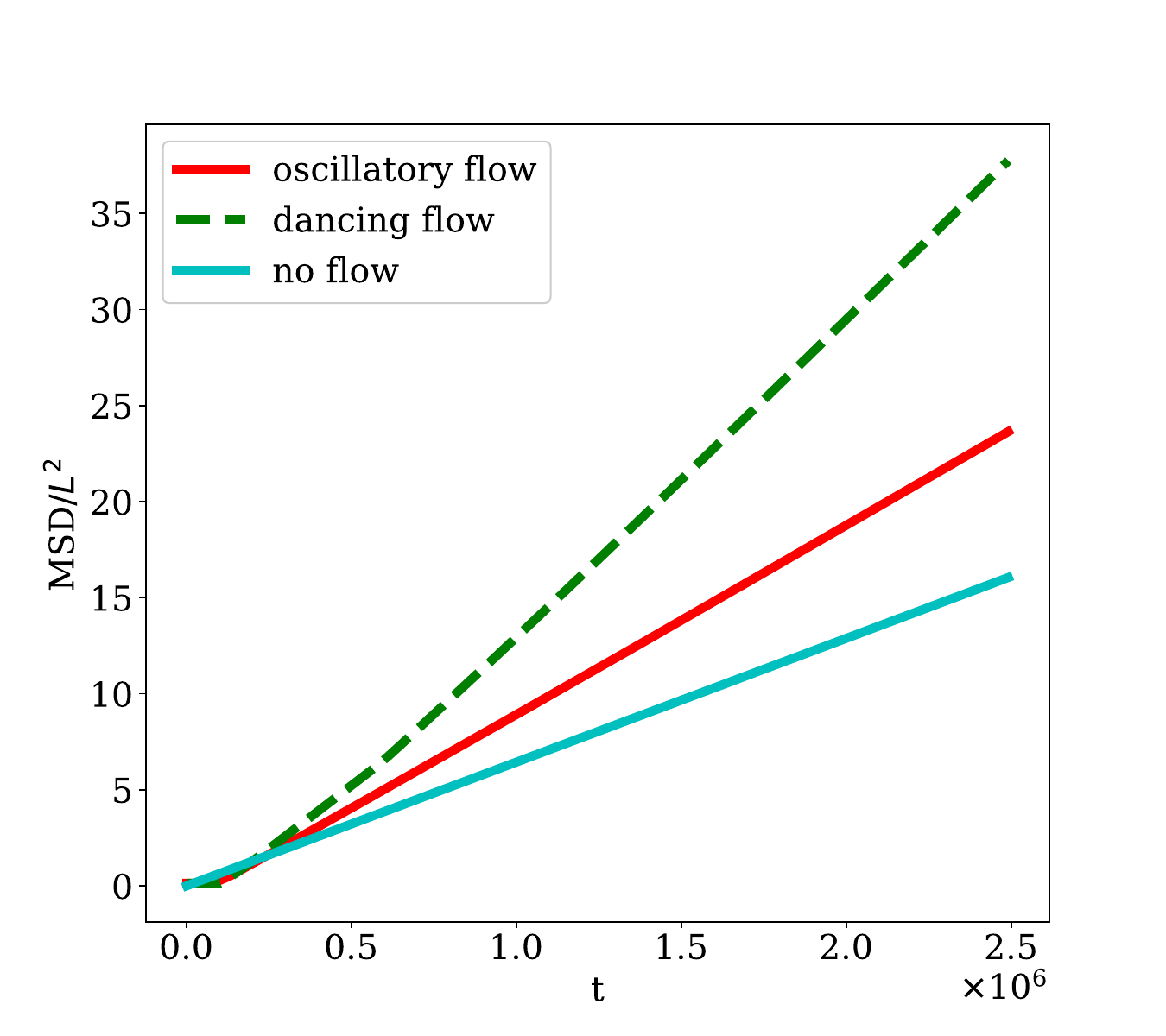}
    \caption{Time evolution of the mean squared displacement (MSD) of solute. The solute MSD is calculated through Eq.~\eqref{eq:msd}. The blue line represents the passive case, with a slope 
     $2D_m$, where $D_m$ is the molecular diffusivity. The red line corresponds to the oscillatory flow, and the green line represents the dancing flow. }
    \label{fig:msdt}
\end{figure}

The system under investigation is a two-dimensional channel filled with an active nematic fluid. The channel is formed by two parallel plates separated by a distance $L=32$. The channel's length $L_x$ is significantly larger, measuring either 1024 or 2048 units, to ensure that the solute's spatial distribution remains unaffected by finite-size effects. No slip boundary conditions for the velocity field at the walls are implemented using an equilibrium scheme \cite{kruger_lattice_2017}. Periodic boundary conditions are applied in the direction parallel to the walls (x-direction). We use planar anchoring only $\mathbf{n}=(1,0)$ with $S=0.5$ at the walls.

We simulate a circular drop of solute placed in the channel after the flow has reached the steady state. The solute's dispersion is quantified by the mean squared displacement (or variance) of its spatial distribution along the channel, calculated from the following expression:
\begin{eqnarray}
    c(x) = \langle c(x,y) \rangle_y = \frac{1}{L}\int_0^{L}c(x,y)dy\\
    \bar{x} = \langle x c(x) \rangle_x \\
    \text{MSD} = \langle c(x) (x-\bar{x})^2 \rangle_x .
    \label{eq:msd}
\end{eqnarray}

\subsection{Dispersion in different flow regimes}

Initially, the system is set with no flow and small uniformly distributed perturbations, up to 2$^\circ$, to
the channel-aligned director field $\mathbf{n}=(1,0)$. The resulting flows are distinguished by a number of features, such as the net mass flux and the dynamics of the director field. In order of increasing activity, we observe the following regimes:

\textbf{Quiescent state} - The velocity field is null everywhere, the director field is uniformly aligned along the channel due to the planar anchoring and the scalar order parameter is $S=0.5$ everywhere. The solute dispersion is purely diffusive. 

\textbf{Defectless vortex lattice} (DVL) - Occurs in the range $\zeta \in [0.005,0.0065]$ or activity number $A \in [18,21]$ where the DVL consists of counter-rotating vortices and a static spatially undulating director field. The DVL was previously described in \cite{gulati_boundaries_2022}, where it was associated to strong anchoring. The time taken to reach this steady state increases with the proximity to the critical activity, separating the quiescent and the DVL regimes. The distortions of the director from the uniformly aligned state are around 5\%. The currents generated by these small gradients do not enhance significantly the dispersion of the solute.

\textbf{Oscillatory flow} - Unlike the vortex lattices that precede and follow it, this regime exhibits net flow. It occurs in the range $\zeta \in [0.0065, 0.011]$ or $A \in [21, 26]$ and is illustrated on the left column of Fig.~\ref{fig: director flow vorticity}.
This type of flow was observed experimentally in \cite{wioland_directed_2016} \cite{hardouin_reconfigurable_2019}  and studied numerically in \cite{chandragiri_active_2019,hardouin_reconfigurable_2019,li_formation_2021,samui_flow_2021,wagner_exploring_2023}. The flow structure, shown in panel~\ref{fig: director flow vorticity} c), is characterized by an undulating region with higher velocity. Examining a constant x-coordinate cross-section, the velocity profile repeats itself in time with a characteristic period T. The ratio of the period to the ``wavelength'' of the undulations (phase velocity) is less than 10 times the average fluid velocity. The nematic order is characterized by two pairs of adjacent topological defects per wavelength, see Fig.~\ref{fig: director flow vorticity} a). There is a symmetric distribution of vorticity, resulting in no net vorticity. 

In Fig.~\ref{fig: soluto} panel a), the fast-moving fluid in the center of the channel pulls the solute away from the slow-moving regions near the walls, as indicated by the orange arrows. This enhances dispersion with a linearly increasing mean squared displacement (MSD), as shown by the red line in Fig.~\ref{fig:msdt}. 
\par 
\textbf{Dancing flow} - The final regime considered here occurs for $\zeta \geq 0.013$ or $A \geq 29$ and forms a vortex lattice with braiding defects. This regime has been extensively simulated and characterized in previous studies \cite{shendruk_dancing_2017, doostmohammadi_onset_2017, coelho_active_2019, hardouin_reconfigurable_2019, chandragiri_active_2019, chandrakar_confinement_2020}, so we refer the reader to these works for further details. In our simulations, this flow regime is shown on the right side of Fig.~\ref{fig: director flow vorticity}. Briefly, the director field, shown in Fig.~\ref{fig: director flow vorticity}b), features quasi-stationary negative topological defects near the channel walls, while positive defects move along the center, avoiding collisions in a manner reminiscent of Ceilidh dancing. The flow structure is characterized by pairs of counter-rotating vortices, as shown in figures~\ref{fig: director flow vorticity} d) and~\ref{fig: director flow vorticity} f). We emphasize that adjacent vortices do not maintain the same absolute vorticity; instead, their relative strengths oscillate over time. Additionally, due to the length of the channel, vortices nucleate at several locations simultaneously, creating asynchronous domains. These domain walls introduce random elements into an otherwise periodic steady state. 

In Fig.~\ref{fig: soluto}b), we see that high and low concentration regions are swept by the vortices, increasing the longitudinal diffusivity by more than 7-fold. 

\begin{figure}
    \centering
    \includegraphics[width=0.4\textwidth]{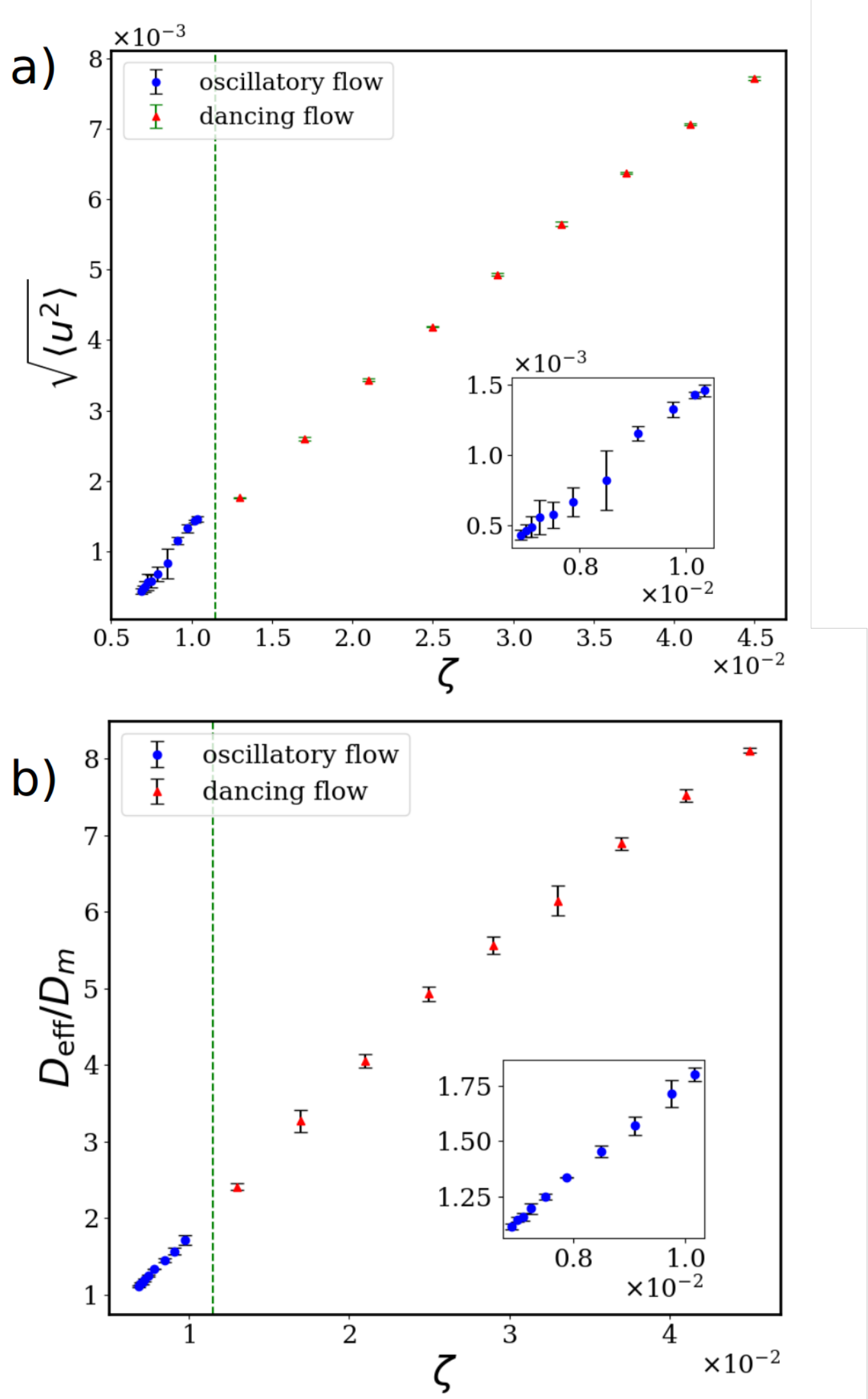}
    \caption{Average velocity of the flow a) and effective diffusivity b) as a function of activity for oscillatory flow and dancing flow. A vertical green line is halfway between $\zeta = 0.010$ and $0.013$, where the transition from oscillatory to dancing flow occurs. The insets show closer views of the oscillatory regime.}
    \label{fig:deff zeta}
\end{figure}

\begin{figure}
    \centering
    \includegraphics[width=0.4\textwidth]{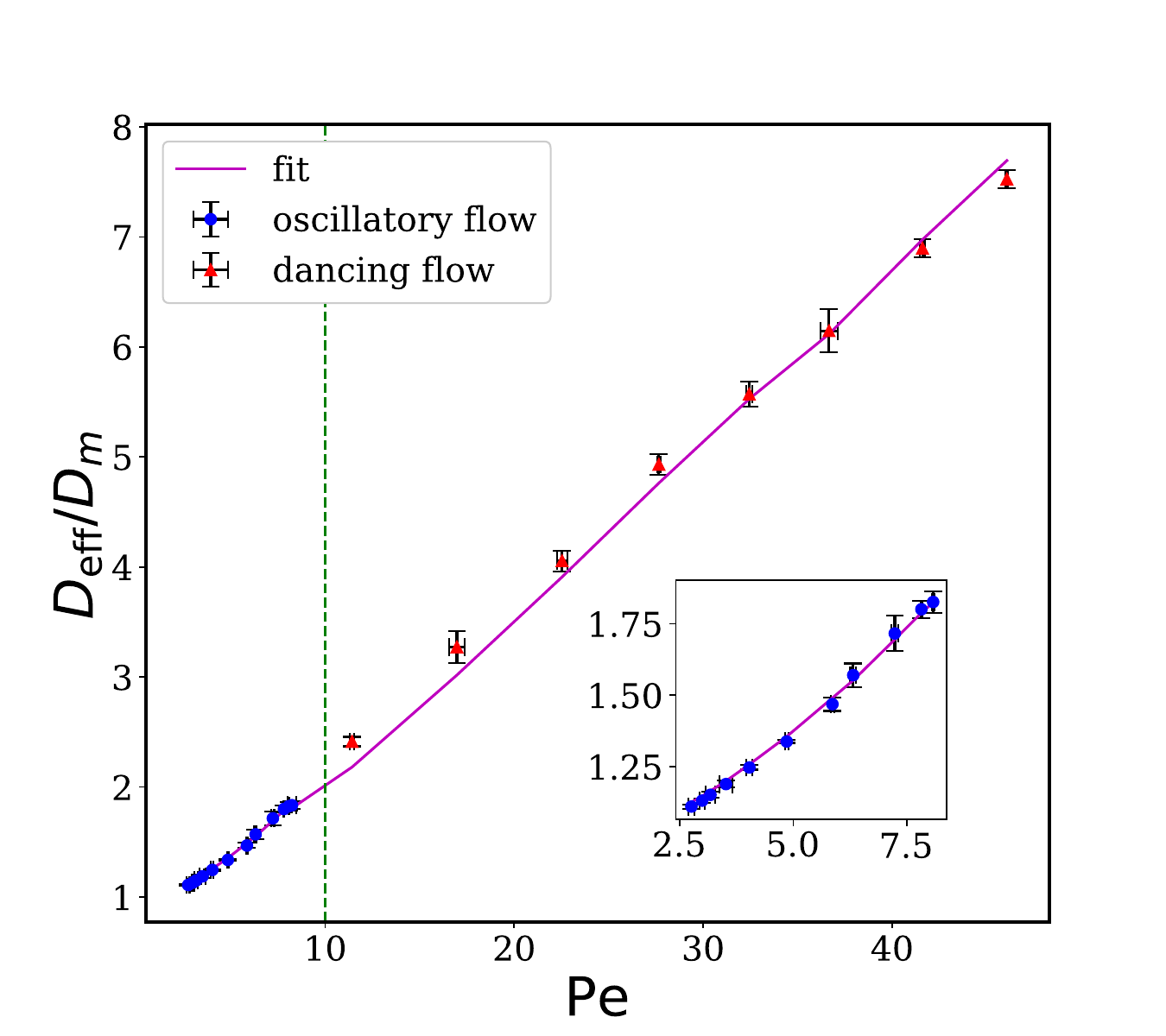}

    \caption{ Effective diffusivity in oscillatory and dancing flow, as a function of the Peclet number $\frac{L}{D_m}\sigma(u_x)$. The points represent the effective diffusivity calculated from the MSD, while the solid line shows the fit from Eq.~\eqref{eq: Active Taylor Dispersion}, with the transverse length $l_{t} = 4.2$ fitted in the oscillatory regime. Note that the fit uses the measured value of $\sigma(u_y)$ for each point. The error bars correspond to the standard deviation calculated from the simulations of 4 samples. The green line, at $\text{Pe}=10$, separates the two regimes.  }
    \label{fig: Active Taylor Dispersion}
\end{figure}

\subsection{Active Taylor-Aris dispersion}

For diffusive dispersion, we can define an effective diffusivity, whose minimum is the molecular diffusivity of the solute (see table~\ref{tab: sim param}), by:
  \begin{eqnarray}
\label{eq: define Deff}
    \text{MSD}(t)= 2 D_\text{eff}t.
\end{eqnarray}
In the oscillatory regime, $D_\text{eff}$ increases linearly with the activity $\zeta$, as shown in Fig.~\ref{fig:deff zeta}b). In the plot of~\ref{fig:deff zeta}a) we find that the characteristic velocity of the flow $v=\sqrt{\langle u^2 \rangle}$ also increases linearly with the activity $\zeta$, by contrast to the sub-linear increase $\zeta^{1/2}$ reported in fully developed active turbulence \cite{hemingway_correlation_2016}. This relationship can be anticipated by dimensional analysis where the channel width $L$ becomes the characteristic length at small activities. In the Stokes regime, Eq.~\eqref{eq: NS} becomes $\eta \nabla^2 \mathbf{u}= \zeta \nabla \cdot \mathbf{Q}$. Thus, the dimensional analysis yields $U \sim \zeta L/\eta$.
Additionally, the linear increase of the flow velocity is in line with the results of the linear stability analysis reported in \cite{voituriez_spontaneous_2005}. However, unlike $D_\text{eff}$, the velocity $v$ increases more rapidly in the oscillatory than in the dancing flow regime. A fraction of the higher energy in the oscillatory flow contributes to the displacement of the cloud of solute center of mass, rather than to enhancing dispersion.

In Poiseuille flow (for passive Newtonian fluids), the solute is spread longitudinally with the width of the solute band increasing with $\sqrt{t}$. The phenomenon is called Taylor-Aris dispersion,  
and it can be characterized by the effective diffusivity:
\begin{eqnarray}
\label{eq: regular TA}
D_\text{eff}= (1 + \frac{\text{\text{Pe}}^2}{210} )D_m,
\end{eqnarray}
where $\text{Pe} = \frac{LU}{D_m}$, $U= \langle u_x \rangle$ is the average fluid velocity, and the coefficient $1/210$ is a geometrical factor for quasi-2d infinite parallel plates \cite{parks_tayloraris_2007}. In Poiseulle flow, the average velocity squared is proportional to the standard deviation of the velocity $U^2 = 5\sigma^2(u_x)$. In the oscillatory flow regime, we found that $\langle u_x \rangle$ and $\sigma(u_x)$ are not proportional. In the dancing flow regime, the average velocity is $\langle u_x \rangle=0$, yet dispersion is also enhanced (Fig.~\ref{fig:deff zeta}). Thus, it is reasonable to assume that the increase in the diffusivity relative to $D_\text{eff}-D_m$ is proportional to the variance $\sigma^2(u_x)$ rather than to the mean velocity squared $\langle u_x\rangle^2 $. 

Replacing $U^2$ by $5\sigma^2(u_x)$ in Eq.~\eqref{eq: regular TA}, however, overestimates the effective diffusivity $D_\text{eff}$ in the active flows considered here, as it neglects the transport of solute from the fast-moving central region of the channel to the slow-moving regions near the walls, and vice versa. This transport may be quantified by the effective diffusivity in the $y$ direction. This should depend on $\sigma(u_y)\sim\sqrt{\langle u_y^2 \rangle}$, in a similar way to the dependence of $D_\text{eff}$ on $\sigma(u_x)$ in the longitudinal ($x$) direction. Therefore, we propose an active form for the Taylor-Aris dispersion, given by:
\begin{eqnarray}
   D_\text{eff}  = D_m + \frac{L^2}{42}\frac{\sigma^2(u_x)}{D_m+l_t \sigma(u_y)}, \label{eq: Active Taylor Dispersion}
\end{eqnarray}
where $l_t$ is a characteristic length, related to the transverse flow, which can be obtained by fitting to the results of numerical simulations. The value of this parameter obtained by fitting Eq.~\eqref{eq: Active Taylor Dispersion} in the oscillatory flow regime was found to be $l_t=4.2$. This expression for the effective diffusivity $D_\text{eff}$ reduces to that of Taylor-Aris dispersion in passive channel flow, Eq.~\eqref{eq: regular TA}, where there is no transverse velocity, i.e.,  $\sigma(u_y)=0$. The fit to the numerical data in the oscillatory flow is shown in the inset of Fig.~\ref{fig: Active Taylor Dispersion}. Note that this fit requires the measured value of $\sigma(u_y)$ for each point. We can rationalize the form of the active Taylor-Aris dispersion in the oscillatory flow regime based on its similarity to Poiseuille flow, from which the original theory was derived. Remarkably, the active Taylor-Aris dispersion describes not only the dispersion of a solute in oscillatory flows but also does so, with the same transverse length $l_t=4.2$, in the dancing flow regime, as shown by the extrapolation depicted in Fig.~\ref{fig: Active Taylor Dispersion}. 
In order to compare the dispersion in oscillatory flows with that in Poiseuille flow with the same flow rate, we confine the standard deviation of the transverse velocity $\sigma(u_y)$ to the range $[3.0\times10^{-4}, 6.5\times10^{-4}]$. The increase in the effective dispersion coefficient, due to the variance $\sigma^2(u_x) $ of the longitudinal velocity in oscillatory flows, is $\frac{D_m}{D_m+l_t \sigma(u_y)}$. This yields an increase in the effective dispersion coefficient that ranges from 2.5 times for the highest activities to 3.8 times for the lowest. 

\begin{figure*}
    \centering
    \includegraphics[width=\textwidth]{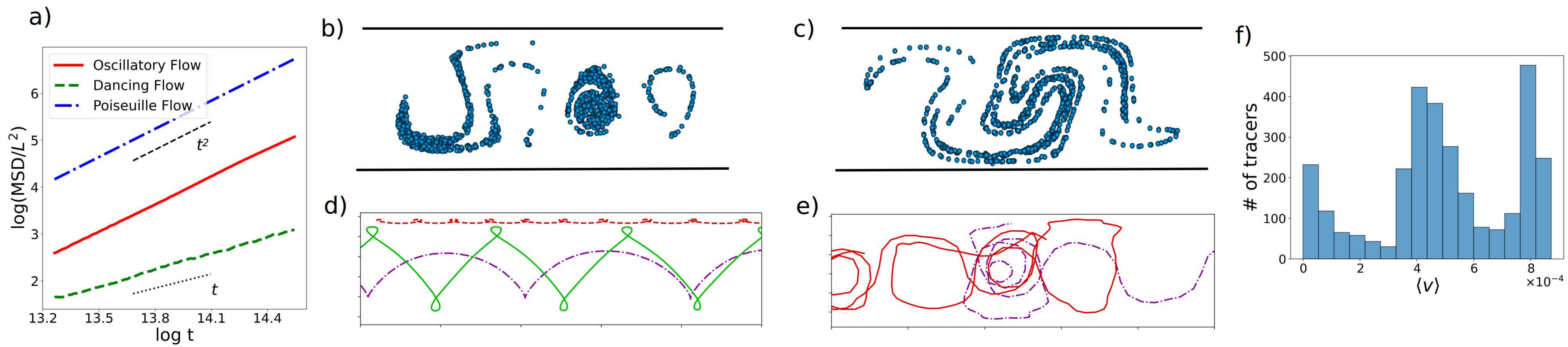}
     \caption{Dynamics of tracers and their trajectories. In a), the log-log plot shows the MSD of tracers, calculated from Eq.~\ref{eq:msdTr}. The initial configuration is 1000 tracers distributed uniformly in a circle of diameter equal to half of the channel’s width. The tracers are released when the flow reaches a steady state. The comparison includes the dispersion in Poiseuille flow with the same flow rate as the oscillatory flow ($\zeta=7.9\times 10^{-3}$). In both oscillatory and Poiseuille flows, the MSD of tracers grows ballistically, while in the dancing flow, the dispersion is diffusive. In b) and c), the tracers are shown shortly after they are released into the oscillatory and dancing flows respectively (see Supplementary Videos 3 and 4). In the panels that follow, the initial configuration is uniform in the whole channel. Panel d) illustrates three typical trajectories in a subsection of the channel in the oscillatory flow regime. In order of increasing average velocity and distance from the walls, the trajectories are red, green, and purple. In e), we illustrate typical trajectories in the dancing flow regime. The histogram in f) depicts the distribution of the average of the absolute value of the x-component of the velocity, defined in the text. The peaks correspond to the 3 different trajectories shown in d). A plot for the dancing flow regime is not shown as the distribution is uniform and the average velocity decreases with time due to the diffusive motion.}
    \label{fig: tracers}
\end{figure*}

\subsection{Convection-dominated limit}

Molecules undergo diffusion driven by thermal fluctuations. However, as the particle size increases, diffusion decreases, as described by the Stokes-Einstein relation $D_m = \frac{k_B T}{6 \pi \eta R}$, for spherical particles of radius $R$, where $k_B$ is the Boltzmann constant and $T$ is the temperature. As the particle diffusivity decreases, the Peclet number, $\text{Pe} = \frac{LU}{D_m}$, diverges for finite $U$. Therefore, large colloids should exhibit much higher $\text{Pe}$ than the solutes depicted in Fig.~\ref{fig: Active Taylor Dispersion}. Active fluctuations, such as those occurring in the dancing flow regime, become more relevant than thermal fluctuations as the driving mechanism for particle dispersion. Additionally, tracers can be easier to study in experiments~\cite{velez-ceron_active_2024} and are useful to follow the trajectories of fluid particles.

In a simplified analysis, we track individual tracers with diffusivity $D_m = 0$. We initialize $N = 1000$ tracers under two distinct conditions: (1) uniformly distributed within a circle of diameter equal to half of the channel width, enabling a direct comparison with the solute dispersion, and (2) uniformly distributed throughout the entire channel, ensuring that trajectories near the walls are included. After a transient regime, tracer $i$ follows the velocity field with $\frac{d \mathbf{x_i}(t)}{dt} = u(\mathbf{x_i}, t)$,
and the cloud's center of mass along the $x$-direction is computed as:$\langle x \rangle = \frac{1}{N} \sum_{i=1}^{N} x_i(t)$,
where $x_i$ is the position of tracer $i$ along the channel. To quantify the dispersion, we compute the mean squared displacement (MSD):
\begin{equation}
    \text{MSD}(t) = \frac{1}{N} \sum_{i=1}^{N} \left[ x_i(t) - \langle x \rangle(t) \right]^2.\label{eq:msdTr}
\end{equation}

The average velocity along the $x$-axis is given by: $\langle v_i \rangle = \frac{\Delta x_i}{\Delta t}$, 
where $\Delta t$ is the total simulation time minus the transient. In the oscillatory regime and for the uniform initial condition, the distribution of $\langle v \rangle $ is tri-modal. The three modes correspond to the trajectories shown in Fig.~\ref{fig: tracers}b. The distribution suggests that once a tracer is set on one of these trajectories, it remains for very long on that trajectory. The crossing of these trajectories implies that the motion depends not only on the distance from the channel walls but also on the initial phase of the traveling wave. Tracers exhibit higher average velocities as they move closer to the center of the channel.

In the oscillatory flow regime, the red curve in Fig.~\ref{fig: tracers}a indicates that tracers spread ballistically, as predicted by the linear stability analysis for unidirectional flow~\cite{basu_anomalous_2012}. While the assumption of $D_m = 0$ is not realistic, especially given the prevalence of fluctuations in pre-turbulent flows, it is safe to assume that the time taken for the MSD to transition from ballistic to diffusive behavior increases with $\text{Pe}$.

In the dancing flow regime, which is closer to turbulence, active fluctuations take the form of perturbations in both the spatial and temporal order, as shown in the array of vortices of Fig.~\ref{fig: director flow vorticity}(d). The local vorticity strength is not 
fully periodic in space and it oscillates irregularly in time. As a result, irregular tracer trajectories are observed as shown in Fig.~\ref{fig: tracers}c. Tracers frequently loop once or twice around a vortex before being ejected to a neighboring vortex. In contrast to the oscillatory regime, tracers in the dancing flow regime exhibit diffusive dispersion, as shown in the log-log plot of Fig.~\ref{fig: tracers}a. It is important to note that the difference between the dispersion of diffusive solute and ballistic tracers is also observed in Poiseuille flow, in line with Eq.~\ref{eq: regular TA}. In active flows, the dispersion of tracers becomes diffusive when the active fluctuations become sufficiently strong to overcome the ballistic transport.

\section{Conclusions}

We investigated the dispersion of solutes in two pre-turbulent flow regimes, resulting from channel confinement of active nematic fluids: oscillatory flow, characterized by net mass flux, and dancing flow, characterized by a one-dimensional vortex lattice, lacking net flux. The results of numerical simulations revealed that the longitudinal dispersion of solutes in both flow regimes is determined by the second moments of the velocity field, specifically the variance of the longitudinal and transverse flows. Furthermore, the linear increase of the dispersion with the activity, irrespective of the presence of vortices or mass flux, reveals a generic or universal picture of dispersion in active nematic flows. The active form of the Taylor-Aris dispersion, which incorporates the effects of transverse flows, applies equally in both flow regimes, with a single fitted parameter. 
 
By contrast, tracers, which do not diffuse due to thermal fluctuations, exhibit distinct dispersion behaviors in the two flow regimes. In oscillatory flows, the tracers may be trapped by vortices near the channel walls or follow high-velocity, tortuous paths, with ballistic dispersion. In the dancing flow regime, however, tracers are flung from vortex to vortex in an irregular manner, resembling one-dimensional random walks, consistent with a diffusive dispersion.\par



Extensions of this work include the study of solutes dispersion in active fluids in porous media, with pores of different sizes and shapes, bringing in a different source of disorder. Additionally, the dispersion of the dissolved chemical energy source may also be considered. Previous studies of the dispersion of activity reported subdiffusive dispersion in the active turbulent regime~\cite{Coelho2022, bate_self-mixing_2022}. It remains to be seen if this result holds in pre-turbulent flow regimes and how it impacts the dispersion of solutes.


Since active turbulence is generic, the flow regimes analyzed here resulting from its confinement, should also be applicable widely, from microtubules to bacteria. The results reported here enhance our understanding of how molecules and colloids, critical to biological processes, are dispersed in confined active fluids in porous media and microfluidic devices.

\section*{Acknowledgements}

We acknowledge financial support from the Portuguese Foundation for Science and Technology (FCT) under the contracts: PTDC/FISMAC/5689/2020 (DOI: 10.54499/PTDCFIS-MAC/5689/2020), \\ UIDB/00618/2020 (DOI: 10.54499/UIDB/00618/2020), \\ UIDP/00618/2020 (DOI: 10.54499/UIDP/00618/2020), \\ DL57/2016/CP1479/CT0057 \\ (DOI: 10.54499 DL57/2016/CP1479/CT0057) \\ and 2023.10412.CPCA.A2.

\bibliographystyle{apsrev4-2}
\bibliography{biblio}

\end{document}